\documentclass[aps,prb,twocolumn,showpacs]{revtex4-1}
\usepackage{epsfig}
\usepackage{dcolumn}
\usepackage{graphicx}
\usepackage{subfigure}
\usepackage{epsf}
\usepackage{soul}
\usepackage{graphicx}
\usepackage{esint}
\usepackage{graphicx}
\begin{document}
\title{Phase diagram and polarization of stable phases of (Ga$_{1-x}$In$_x$)$_2$O$_3$}
\author{Maria Barbara Maccioni and Vincenzo Fiorentini}
\address{Department of Physics, University of Cagliari and CNR-IOM, UOS Cagliari, Cittadella Universitaria, 09042 Monserrato (CA), Italy }
\begin{abstract}
 Using  density-functional ab initio  calculations, we provide a revised  phase diagram of  (Ga$_{1-x}$In$_{x})_2$O$_3$. Three phases --monoclinic, hexagonal, cubic bixbyite-- compete for the ground state.  In particular, in the $x$$\sim$0.5 region we expect coexistence of hexagonal, $\beta$, and bixbyite (the latter  separating into binary components). Over the whole $x$ range, mixing occurs in three disconnected regions,  and non-mixing in two additional  distinct regions.
 We then explore  the permanent polarization of the various phases, finding that none of them is polar at any concentration, despite the possible symmetry reductions induced by alloying. On the other hand, we find that the $\varepsilon$ phase of Ga$_2$O$_3$ stabilized in recent growth experiments is pyroelectric --i.e. locked in a non-switchable polarized structure-- with ferroelectric-grade polarization and respectable piezoelectric coupling. We suggest that this phase could be used profitably to produce high-density electron gases in transistor structures.  \end{abstract}
\maketitle

\section{Introduction}
The Ga and In sesquioxides have recently been under intense scrutiny  as, among others, UV absorbers and  trasparent conductors. With a view at exploiting materials engineering concepts from other semiconductor systems, a growing body of work is being devoted to the  (Ga$_{1-x}$In$_x$)$_2$O$_3$ alloy. Theoretical studies on its phase stability and optical properties have been published recently by at least two groups \cite{noi1,noi2,noi3,noi4,janotti}, but the picture is apparently still far from complete. Recent growth experiments \cite{ikz,japepsilon} on the (Ga$_{1-x}$In$_x$)$_2$O$_3$ alloy in the vicinity of $x$=0 and $x$=0.5 have suggested that phases other than those so far assumed as ground state may in fact be stable or stabilized by constrained (e.g. epitaxial) growth. 
In one paper \cite{ikz}, three competing phases are reported to appear near $x$$\sim$0.5:  a hexagonal phase previously observed at exactly $x$=0.5;  a monoclinic close relative of the $\beta$-Ga$_2$O$_3$ structure; and that derived from the bixbyite structure of In$_2$O$_3$, mostly in the form of phase-separated In and Ga oxides. Another paper \cite{japepsilon} reported that the $\varepsilon$ phase of Ga$_2$O$_3$ can be obtained at 820 K via epitaxial growth on GaN, although a  bulk phase transition to that phase from the ground state $\beta$ phase is not expected below 1500 K  \cite{teoriaga2o3}.  

In this work, we report $i$) a   phase diagram of 
(Ga$_{1-x}$In$_x$)$_2$O$_3$ accounting for new findings around $x$$\simeq$0.5, and $ii$) the  polarization of the competing phases, plus the $\varepsilon$-Ga$_2$O$_3$ phase. The results are in a nutshell that  $a$) the hex and $\beta$ structures do indeed compete energetically with the bixbyite phase  expected based on previous results, and this competition occurs predominantly in the vicinity of $x$$\sim$0.5, giving rise to a fairly complex phase diagram with interlacing mixing and non-mixing concentration regions; $b$) none of the alloy phases is polar, but the $\varepsilon$ phase of Ga$_2$O$_3$ is. As dictated by its symmetry, this phase (only slightly  energy-disfavored over the stable $\beta$ phase) has a large spontaneous polarization and sizable piezoelectric coupling. Importantly, it cannot be transformed into (is not symmetry-related to) the stable $\beta$ phase. These results  open up some interesting perspectives, such as growing the hex phase epitaxially, or  exploiting the polarization properties of 
$\varepsilon$-Ga$_2$O$_3$.

\section{Methods}
Geometry and volume optimizations as well as electronic structure calculations are performed using density-functional theory (DFT) in the generalized gradient approximation (GGA), and the Projector Augmented-Wave (PAW) method as implemented in the VASP code \cite{vasp}. In all calculations the cut-off is 471 eV and the force threshold is 0.01 eV/\AA. For all phases, 80-atom cells are used. The k-point summation grids are a $\Gamma$-centered  2$\times$2$\times$2  for the hexagonal phase and 4$\times$4$\times$2 for the $\varepsilon$ phase (4$\times$4$\times$2 for the $\beta$ phase and 4$\times$4$\times$4 for the bixbyite structure as in Ref.\onlinecite{noi3}). The polarization is calculated via the Berry-phase approach;\cite{berry} the k-grid is a 4$\times$4 set of 16-point strings. 

Phase coexistence is determined by the Helmholtz mixing free energy per  cation (the enthalpy  vanishes because the pressure is kept to  zero to numerical accuracy in all calculations)  of the mixture as a function of $x$,
\begin{equation}
F_{\rm mix} = F_{\rm alloy} - F_{\rm bulk}=
[E_{\rm alloy} - T S_{\rm alloy}] - F_{\rm bulk},
\end{equation}
where
$E_{\rm alloy}$ is the internal energy calculated  from first principles as just described, and
\begin{eqnarray}
S_{\rm alloy}&=&
 -x\, \log{x} - (1-x)\, \log{(1-x)} +\\& &+ 3\,\left[(1+n)\, \log{(1+n)}-n\, \log{n}\right] = S_{\rm mix}+S_{\rm vib}\nonumber
\end{eqnarray}
with 
\begin{eqnarray}
n (T,x) &=& 1/(e^{\Theta_m(x)/T}-1),\nonumber \\
\Theta_m (x) & =& (1-x)\, \Theta_{\rm Ga_2O_3} + x\, \Theta_{\rm In_2O_3}
\end{eqnarray}
the Planck distribution and the mixture's  Debye temperature $\Theta_m$($x$)  interpolated between the parent compounds.  (The approximation of the vibrational entropy with that of a single-Debye-frequency oscillator is admissible, as the growth  temperatures are comparable to or higher than the Debye temperatures $\Theta_{\rm In_2O_3}$=420 K and $\Theta_{\rm Ga_2O_3}$=870 K.)
The bulk free energy 
\begin{equation}F_{\rm bulk} (x) = x\, F_{\rm In_2O_3} + (1-x)\, F_{\rm Ga_2O_3}
\end{equation}
interpolates between the binary-compound values, calculated as for the alloy.
We finally recall that a mixture separates into phases if the  specific free energy is a negative-curvature function of  $x$. The $x$ values where the curvature becomes negative and, respectively, goes back to positive (i.e. the inflection points of the mixing free energy) delimit the phase separation region; these bounds, which may depend on temperature T,  define a range known as  miscibility gap.  
\section{Results}
\subsection{Hexagonal phase near $x$$\simeq$0.5}

\begin{figure}[ht]
\centering
\includegraphics[width=9cm]{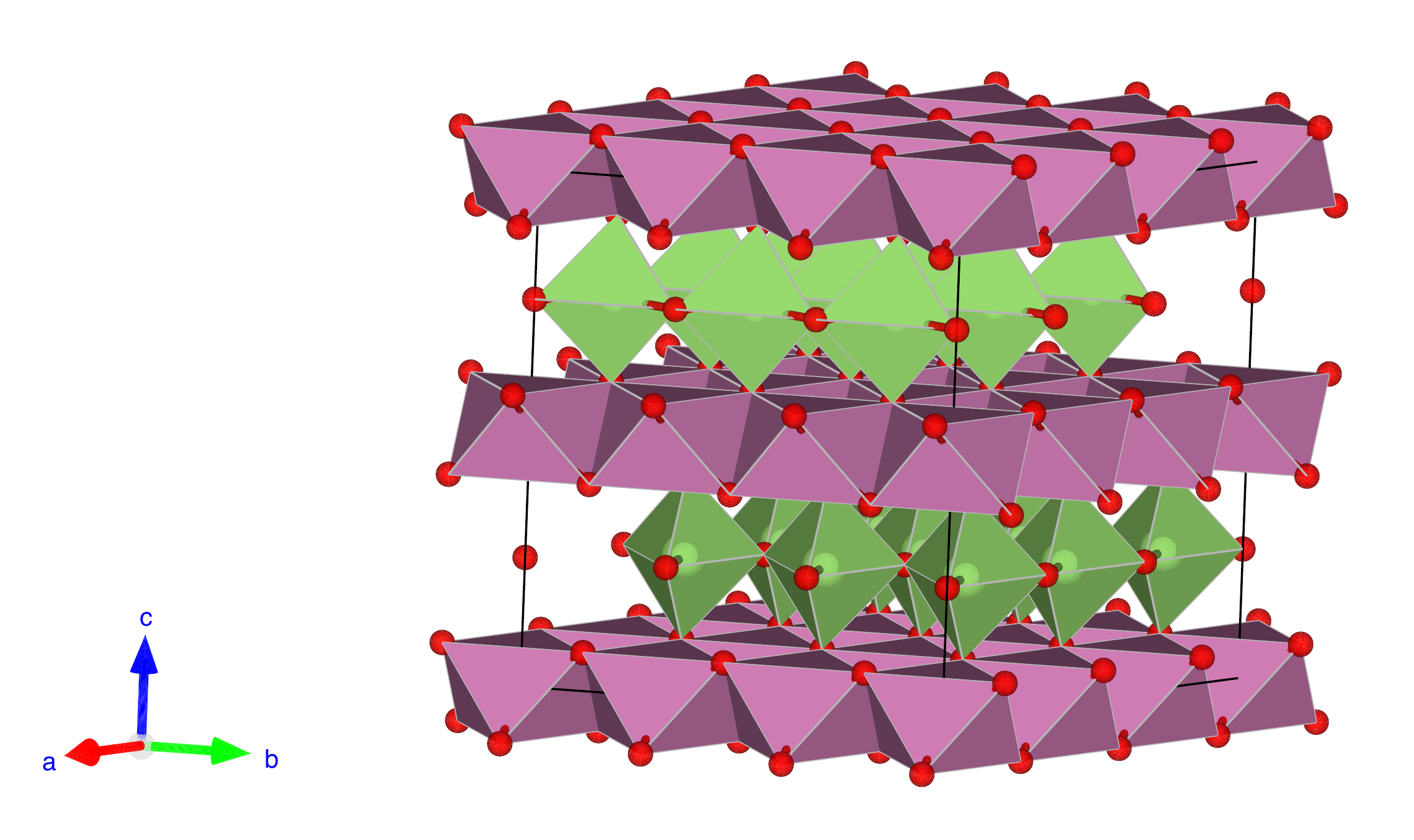}
\caption{\label{hex} The hex structure at 50-50 concentration.}
\end{figure}

We first consider a hexagonal phase of (Ga$_{1-x}$In$_x$)$_2$O$_3$. The  motivation comes from recent growth experiments  \cite{ikz} of (Ga$_{1-x}$In$_x$)$_2$O$_3$ near $x$$\sim$0.5, which revealed the appearance, besides $\beta$ and bixbyite crystal portions,  of significant  hexagonal microcrystallites. A candidate phase had been identified earlier on \cite{ingao3}  for InGaO$_3$, and   classified in the non-polar space group $P6_3$/$mmc$. This phase is depicted at $x$=0.5 in Fig.\ref{hex}.

Both out of interest for its possible energetic stability, and for the possibility that the structure might become polar, we investigated this phase in the range $x$$\in$[0.45,0.55].  In this region the hex phase is lower in energy than, and therefore favored over, the bixbyite and virtually degenerate at $x$=0.5 with the $\beta$ phase (discussed below).  We quantify this calculating the Helmholtz mixing free energy by the model described above and in Ref.\onlinecite{noi3}. (As in previous work,\cite{noi1,noi2,noi3} large error bars are expected from the limited  configurational sampling, but as errors should largely cancel out when comparing the various phases, we deem the relative energetics to be rather reliable.)  As shown in Fig.\ref{phdiag}, bottom panel, we find that the lowest energy structures of the sample of configurations for the hex symmetry in the vicinity of $x$=0.5 are lower in mixing free energy by about 0.1 eV than the free-standing bixbyite configurational sample, and therefore more stable than the bixbyite alloy. 

\subsection{$\beta$ phase near $x$$\sim$0.5}

\begin{figure}[ht]
\centering
\includegraphics[width=9cm]{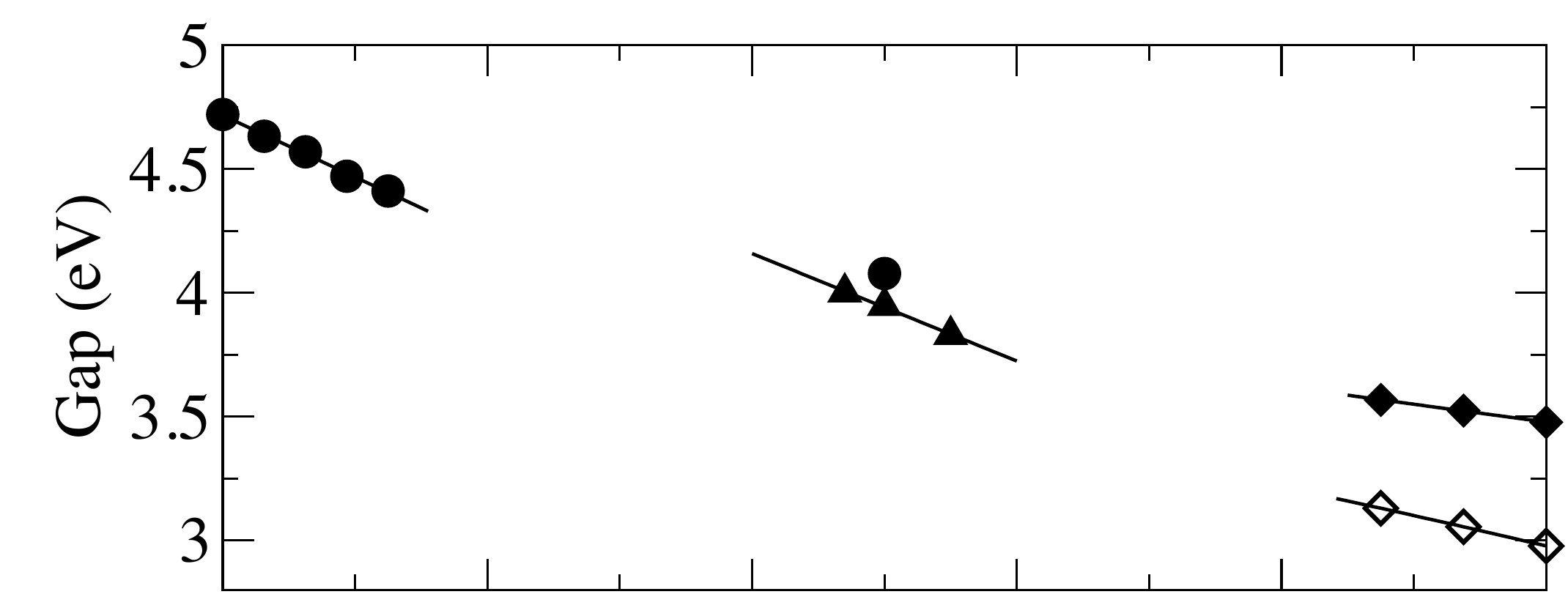}
\includegraphics[width=9cm]{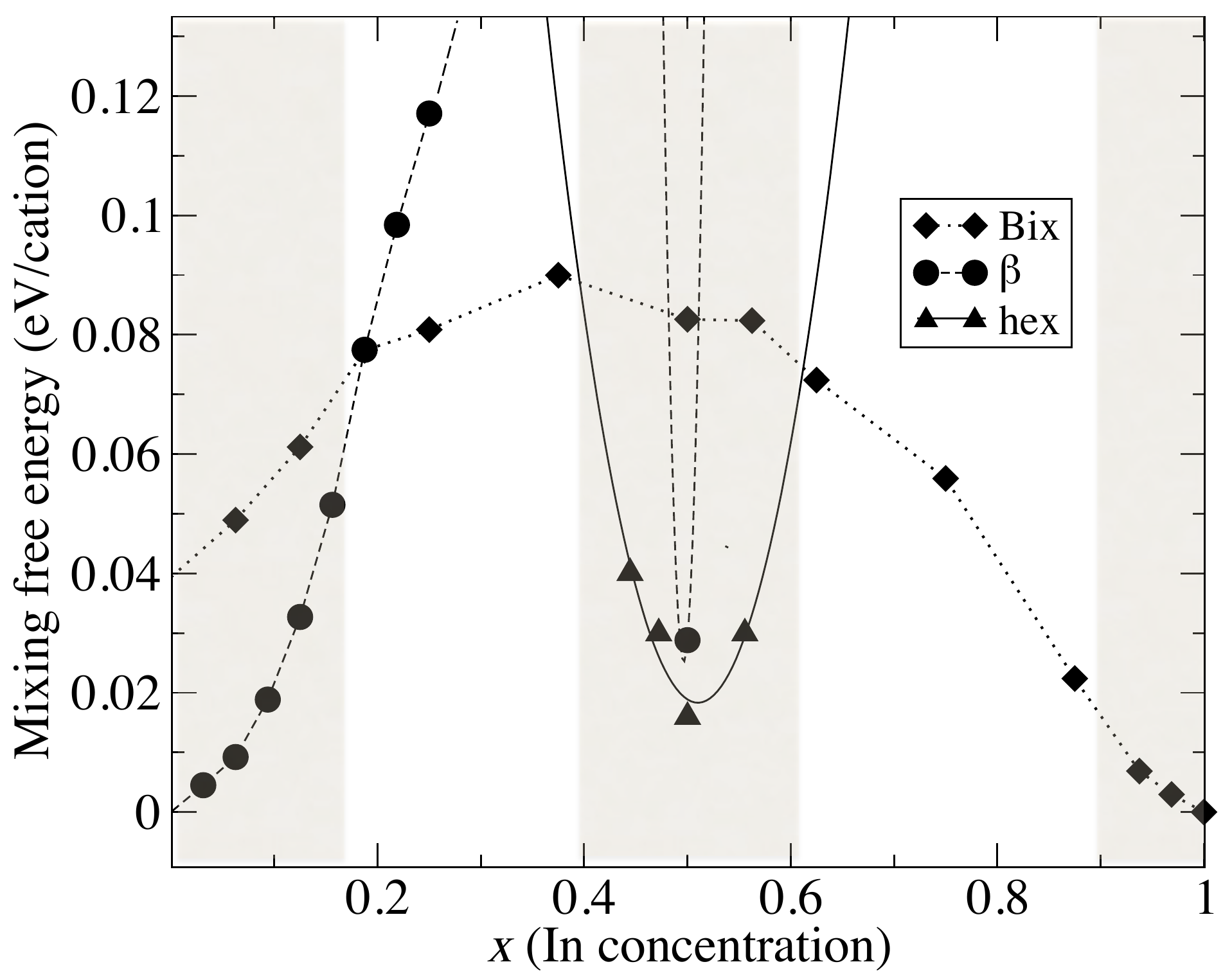}
\caption{\label{phdiag} Revised phase diagram at T=800 K (bottom) and energy gap (top) at T=0 for (Ga$_{1-x}$In$_x$)$_2$O$_3$. 
Mixing regions are shaded, and the gap is only drawn for those regions. Legend applies to both panels; empty diamonds: dipole-forbidden transition in bixbyite. The lines are quadratic fits for hex and $x$$\sim$0.5, guides to the eye for bixbyite and low-$x$ $\beta$.}
\end{figure}

In previous work \cite{noi1,noi2,noi3} we found that the alloy adopting the  $\beta$ structure of Ga$_2$O$_3$ is disfavored over bixbyite for $x$ above 0.1 or so. The internal energy of that phase increases drastically and monotonically in that region of $x$, so we refrained from pursuing it further. However,   the same paper reporting the occurrence of hex phase crystallites also signaled $\beta$-phase inclusions  near $x$=0.5, so we revisited our previous assessment and studied the $\beta$ phase in that region of concentration. It turns out that at exactly $x$=0.5 the $\beta$ phase is more stable than bixbyite and as stable as the hex phase discussed above (see Fig.\ref{phdiag}, bottom; a similar occurrence was reported in Ref.\onlinecite{janotti}). At this concentration, In atoms  occupy all the octahedral sites, and Ga atoms  occupy all the tetrahedral sites. However, consistently with our previous conclusions, as soon as we move away from exact 50-50 concentration the energy shoots up immediately on both sides of the $x$=0.5 minimum, accompanied by a volume collapse (mainly of the tetrahedra) by   over 10\% at $x$=0.47 and $x$=0.53. Therefore, the $\beta$ phase itself should only occur at the ``magic'' 50-50 concentration, or in the vicinity of that concentration if one assumes  that some other phase  will take up the local cation excess.

\subsection{Revised phase diagram}

Based on our calculations discussed above we provide an improved phase diagram  accounting for the new phases. The diagram is reported in Fig.\ref{phdiag}, bottom panel, as mixing free energy vs $x$. The temperature is 800 K, a typical growth temperature. 
As shown previously,\cite{noi3} the phase boundaries are insensitive to temperature within our model, and hence apply to all practical growth temperatures. Put differently, the miscibility gaps and miscibility regions are persistent with temperature.

The stability of the $\beta$ phase only at low $x$ is confirmed, and so is  the phase separation into components of the bixbyite phase in most of its own range (signaled by the free energy being everywhere upward-convex except for $x$$\geq$0.9). What is new is that the hex phase is now the stable one in a range that, conservatively, extends from $x$$\sim$0.4 to $x$$\sim$0.6. Given its upward-concave free energy, the hex phase does not phase-separate into binary components in this range. In addition, as mentioned, the $\beta$ phase has a very narrow stability slot at $x$=0.5. 

As dictated by the curvature of the mixing free energy,  there is full miscibility of the two binary oxides at all temperatures in the ranges $x$$\in$(0,0.18), $x$$\in$(0.4,0.6), and
 $x$$\in$(0.9,1), where, respectively, the $\beta$, the hex, and the bixbyite structures are adopted.   (These $x$ values correspond to the inflection points of the mixing free energy.) In the rest of the $x$ range,  separation into binaries is expected from the convex mixing free energy.  
 
In the central region of the $x$ range, there are several competing possibilities. The hex and $\beta$ mixed phases are obviously favored over the bixbyite alloy. But the latter  should phase-separate into binary components, with In$_2$O$_3$ certainly adopting the bixbyite structure. Ga$_2$O$_3$ may go either  bixbyite or $\beta$: in the first case,  the energy (from an interpolation between the end values) is about 0.02 eV, i.e.  falls between the $\beta$ alloy and the  hex; in the second case, the free energy is zero by construction, making phase separation at 50\% slightly favored. These considerations, however, neglect internal interfaces, grain boundaries, strain effects in the binaries, and growth kinetics (all of which are  exceedingly complicated and well beyond our present scope), which will tend to disfavor phase separation. Thus, at this level of accuracy,  it seems very plausible that --as experiments suggest-- the hex, $\beta$, and  phase-separated binaries  will coexist in this region, depending on the growth conditions. 
  
In Fig.\ref{phdiag}, top panel, we also report the calculated fundamental gap in the stable phases in the regions where mixing occurs. The gap is calculated as difference of Kohn-Sham GGA eigenvalues plus  the empirical correction used in Ref.\onlinecite{noi3} to adjust the gap of the binaries to the experimental values. At low $x$ and at high $x$ one expects optical absorption typical of Ga$_2$O$_3$ and In$_2$O$_3$, respectively. Around $x$$\sim$0.5, the hex and $\beta$ alloy absorptions should be present; since the bixbyite alloy phase separates, absorption may also be observed at energies typical of  Ga$_2$O$_3$ (4.5-4.7 eV) and In$_2$O$_3$ (2.9 eV forbidden, 3.5 eV allowed)); so in the central $x$ region, distinct transitions may be expected at roughly 3.5 eV, 4 eV and 4.5 eV.

\subsection{Polarization: $\varepsilon$-Ga$_2$O$_3$}
 
One  reason of interest in the hex phase is checking whether it  distorts into a non-centrosymmetric symmetry group  as a consequence of alloying. We investigate a range of alloying of between 43\% and 57\%, enabling all symmetry lowerings starting from  $P6_3$/$mmc$. We  find that the polarization is always numerically zero referred to the non-polar high-symmetry phase, and are thus forced to conclude that this structure, somewhat anticlimatically, is robustly non polar. In fact, our conclusion agrees with the symmetry determination of Ref.\onlinecite{inorgch} at $x$=0.5, and shows that this applies at generic concentrations in that vicinity. We also sampled the polarization in a few bixbyite and $\beta$ alloy samples, unsurprisingly finding them to be always zero (referred to the cubic and monoclinic-$\beta$ phases). 

We therefore turned to another potentially polar phase of this system,  $\varepsilon$-Ga$_2$O$_3$,  recently grown epitaxially on GaN by  Oshima {\it et al.} \cite{japepsilon}.
The $\varepsilon$ phase is structurally akin to the same phase of Fe$_2$O$_3$ and its space group  is $Pna2_1$, which does not contain inversion. We calculate its structural parameters, finding them in essential agreement with a previous study, \cite{teoriaga2o3} and the energy difference with the $\beta$ phase at zero temperature, which is just 90 meV per formula unit. 

At low enough temperature, the epitaxial sta\-bi\-li\-za\-tion \cite{japepsilon} of the $\varepsilon$ phase is not endangered by a possible decay in the $\beta$ ground state, for the simple reason that there is no possible
$\varepsilon$-to-$\beta$ symmetry path, since the two space groups are $Pna2_1$ and  $C2/m$, respectively.  This is quite analogous to the situation of wurtzite III-V nitrides (group $P6_3mc$) which cannot transform, again for symmetry reasons, into the close relative structure of zincblende (group $F\overline{4}3m$), despite the volume being almost the same and the energy difference being about only 10 meV/atom (the two $\varepsilon$ and $\beta$ phases also have the same volume and an energy difference of about 15 meV/atom). Just as  $\varepsilon$-Ga$_2$O$_3$, zincblende nitrides can be grown under appropriate epitaxial constraints.

\begin{figure}[ht]
\centering
\includegraphics[width=9cm]{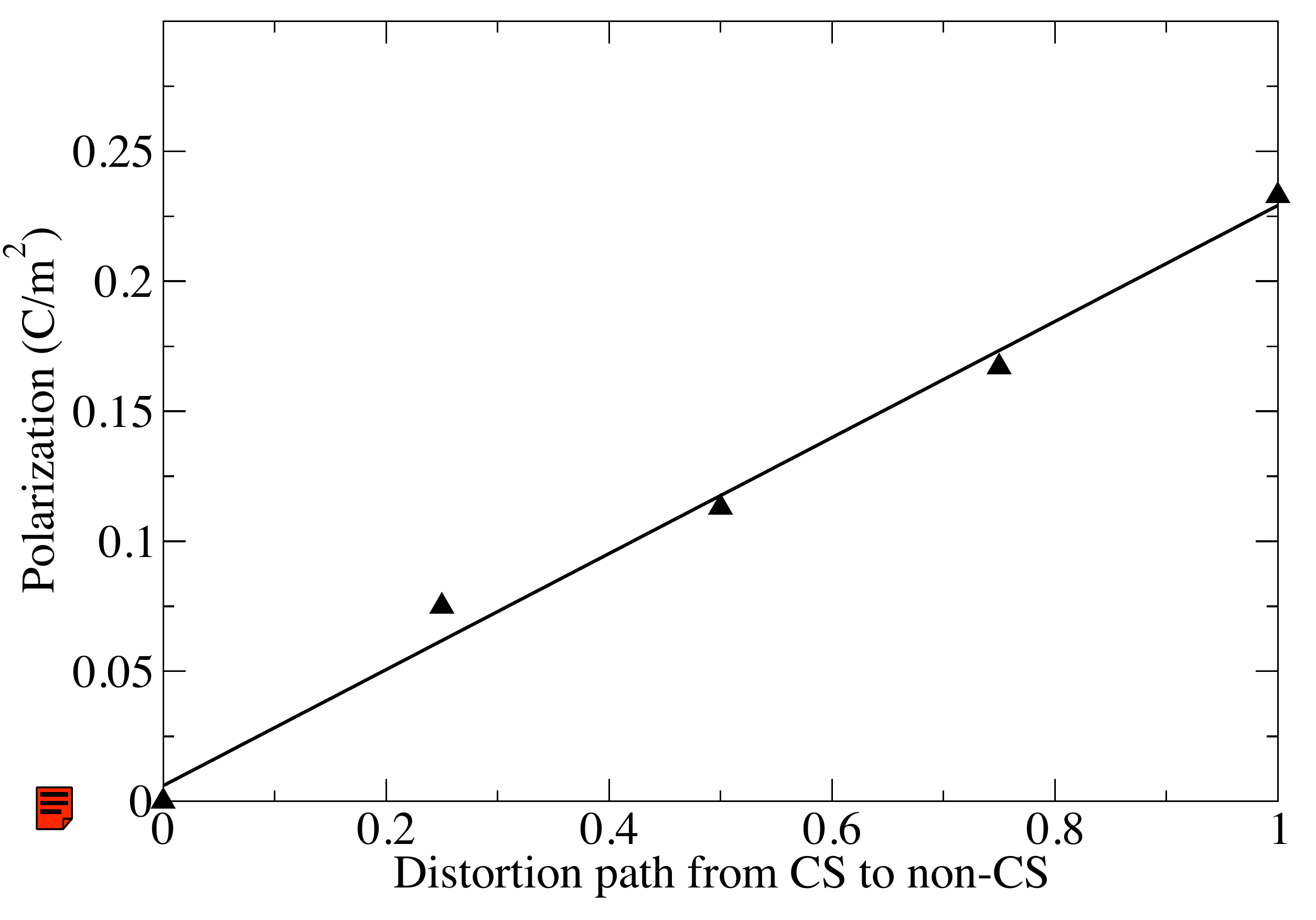}
\caption{\label{pvsdis} Polarization calculated along a path connecting the centrosymmetric parent phase to non-centrosymmetric $\varepsilon$-Ga$_2$O.}
\end{figure}

Since the  group $Pna2_1$ does not contain inversion, observable polar vector quantities are allowed in  $\varepsilon$-Ga$_2$O$_3$. The polar axis is the $c$ axis, so we calculate the spontaneous polarization {\bf P}=(0,0,P)  as difference of the   polarizations calculated  \cite{berry} in this phase and in a symmetry-connected centrosymmetric parent phase (we verified that the other components are indeed zero). The latter is chosen to have symmetry group $Pnma$ (a supergroup of  $Pna2_1$). The evolution of P with a path connecting the two structures is in Fig.\ref{pvsdis}.

The final result is that $\varepsilon$-Ga$_2$O$_3$ has a remarkable P=0.23 C/m$^2$, a value similar to that of BaTiO$_3$, a factor of 3 larger than of AlN, and nearly a factor 10 larger than of GaN. The structure of the $\varepsilon$ phase is not structurally switchable (in the same sense that wurtzite is not, though of course the polar axis can be inverted, again as in nitrides, by inverting the growth direction using e.g. a buffer layer); therefore P is expected to mantain its orientation along the polar axis within any given crystalline domain.  $\varepsilon$-Ga$_2$O$_3$ can  thus be classified as a pyroelectric material, one more time like III-V nitrides. The symmetry of the $\varepsilon$-Ga$_2$O$_3$ structure allows for five distinct piezoelectric coefficients; here we calculate the diagonal coefficient $e_{33}$ as the finite-differences derivative of the polarization with respect to the axial strain $\epsilon_3$=($c$--$c_0$)/$c_0$. The result is $e_{33}$=0.77 C/m$^2$, which is in line with typical coefficients of strongly polar semiconductors (oxides and nitrides), although over an order of magnitude  smaller than those of strong ferroelectrics (see e.g.  Ref.\onlinecite{cinesi}).

These results   open up  interesting perspectives. The polarization  of 
$\varepsilon$-Ga$_2$O$_3$ can be exploited growing the oxide epitaxially on GaN (or, equivalently, growing GaN on the oxide) to build a high-mobility transistor. The polar axis of $\varepsilon$-Ga$_2$O$_3$ is found to be parallel to that of GaN in growth experiments,\cite{japepsilon} so their polarizations are either parallel or antiparallel. Since the polarization difference is very large in both cases,  a correspondingly large polarization charge will appear at the GaN/$\varepsilon$-Ga$_2$O$_3$ interface and thus attract free carriers (provided by dopants, e.g.) to form an interface-localized two-dimensional gas at potentially huge concentrations. The polarization difference across the interface, i.e. the polarization charge to be screened by free carriers, and hence the potentially-reachable local electron-gas concentration, is 0.2 to 0.26 C/m$^2$, i.e. 1.2 to 1.6$\times$10$^{14}$ cm$^{-2}$ (for  {\bf P} vectors in the two materials being parallel or antiparallel, and neglecting possible interface traps, native charges, etc.). In addition, since the gap of $\varepsilon$-Ga$_2$O$_3$ is much larger than that of GaN, the interface confinement should be quite efficient. Finally, the sign of the polarization charge and hence of the accumulation layer will depend on the chosen polarity of the substrate. The above scenario is  a ``writ-large'' version of the GaN/AlGaN HEMTs currently in use, whose  high-frequency, high-power operation is enabled primarily by the high polarization-induced interface charge. Of course,  all else assumed to be equal, Ga$_2$O$_3$/GaN transistors could be much superior to AlGaN/GaN ones, which enjoy a
much lower areal density of order 10$^{13}$ cm$^{-2}$.
 
\section{Summary}

Using  density-functional ab initio  theoretical techniques, we have revised the phase diagram of   (Ga$_{1-x}$In$_x$)$_2$O$_3$, showing that  the $\beta$ phase is stable (without phase separation into binary components)  at low $x$ and exactly at 50-50 concentration; a new hexagonal phase is stable (again without phase separation into binary components) for $x$ from about 0.4 to 0.6, where it is robustly non-polar;  and bixbyite will be favored for $x$ between 0.2 and 0.4 and upward of 0.6, but should phase-separate into binary components. Around $x$$\sim$0.5, the hex, $\beta$ and phase-separated binary bixbyites should be closely competing. Optical signatures are expected at around 4.6 eV at low $x$ ($\beta$ phase), around 3.5 eV at large $x$ (bixbyite), and at 3.5, 4, 4.5 eV from the competing phases at  $x$$\sim$0.5.

We have further studied the $\epsilon$-phase of Ga$_2$O$_3$, and confirmed it as the second most stable structure beside $\beta$-Ga$_2$O$_3$. We find it  to have a large spontaneous polarization (0.23 C/m$^2$) and a sizable diagonal piezoelectric coefficient ($e_{33}$=0.77 C/m$^2$). Symmetry dictates that this phase, once epitaxially stabilized, will not transform back into the ground-state $\beta$, despite having the same volume and a small energy difference; in this sense, the $\varepsilon$-$\beta$ relation is similar to that between zincblende and wurtzite III-V nitrides.

\section*{Acknowledgments}
Work  supported in part by MIUR-PRIN 2010 project {\it Oxide}, CAR and PRID of University of Cagliari, Fondazione Banco di Sardegna grants, CINECA computing grants.
MBM acknowledges  the financial support of her PhD scholarship by Sardinia Regional Government under P.O.R. Sardegna F.S.E. Operational Programme of the Autonomous Region of Sardinia, European Social Fund 2007-2013 -
Axis IV Human Resources, Objective l.3, Line of Activity l.3.1.

\end{document}